\documentclass[a4paper,fleqn,usenatbib]{mnras}

\usepackage{gensymb}
\usepackage{newtxtext,newtxmath}
\usepackage{enumerate}
\usepackage{float}

\def\ha{H$\alpha$}
\def\hb{H$\beta$}

\def\ot{[O\,{\sc iii}]}

\def\com{}

\usepackage[T1]{fontenc}
\usepackage{ae,aecompl}

\usepackage{graphicx}	
\usepackage{amsmath}	
\usepackage{amssymb}	

\DeclareRobustCommand{\De}[3]{#2}

\title[Clusters Over Cosmic Time]{Towards Studying Hierarchical Assembly in Real Time: A Milky Way Progenitor Galaxy at $z = 2.36$ under the Microscope}
    
\author[T. O. Zick]{Tom O. Zick$^{1}$,\thanks{E-mail: tzick@berkeley.edu}
Daniel R. Weisz$^{1}$, 
Bruno Ribeiro$^{2}$, 
Mariska T. Kriek$^{1}$,
Benjamin D. Johnson$^{3}$, 
\newauthor
Xiangcheng Ma$^{1}$, 
Rychard Bouwens$^{2}$
\\
$^{1}$Astronomy Department, University of California, Berkeley, CA 94720, USA\\
$^{2}$Leiden Observatory, Leiden University, NL-2300 RA Leiden, Netherlands\\
$^{3}$Center for Astrophysics, Harvard \& Smithsonian, Cambridge, MA 02138, USA\\}

\date{Accepted XXX. Received YYY; in original form ZZZ}

\pubyear{2019}

\begin{document}
\label{firstpage}
\pagerange{\pageref{firstpage}--\pageref{lastpage}}
\maketitle

\begin{abstract}
We use Hubble Space Telescope (HST) imaging and near-infrared spectroscopy from Keck/MOSFIRE to study the sub-structure around the progenitor of a Milky Way-mass galaxy in the Hubble Frontier Fields (HFF).  Specifically, we study an $r_e = 40^{+70}_{-30}$pc, $M_{\star} \sim 10^{8.2} M_{\odot}$ rest-frame ultra-violet luminous ``clump'' at a projected distance of $\sim$100~pc from a $M_{\star} \sim 10^{9.8}$M$_{\odot}$ galaxy at $z = 2.36$ with a magnification $\mu = 5.21$.  We measure the star formation history of the clump and galaxy by jointly modeling the broadband spectral energy distribution from HST photometry and H$\alpha$ from MOSFIRE spectroscopy.  Given our inferred properties (e.g., mass, metallicity, dust) of the clump and galaxy, we explore scenarios in which the clump formed \emph{in-situ} (e.g., a star forming complex) or \emph{ex-situ} (e.g., a dwarf galaxy being accreted).  If it formed \emph{in-situ}, we conclude that the clump is likely a single entity as opposed to a aggregation of smaller star clusters, making it one of the most dense star clusters cataloged.  If it formed \emph{ex-situ}, then we are witnessing an accretion event with a 1:40 stellar mass ratio.  However, our data alone are not informative enough to distinguish between \emph{in-situ} and \emph{ex-situ} scenarios to a high level of significance.  We posit that the addition of high-fidelity metallicity information, such as [O\,{\sc iii}]4363\AA, which can be detected at modest S/N  with only a few hours of JWST/NIRSpec time, may be a powerful discriminant.  We suggest that studying larger samples of moderately lensed sub-structures across cosmic time can provide unique insight into the hierarchical formation of galaxies like the Milky Way.

\end{abstract}




\section{Introduction}
\begin{figure*}
\includegraphics[width=6.8in]{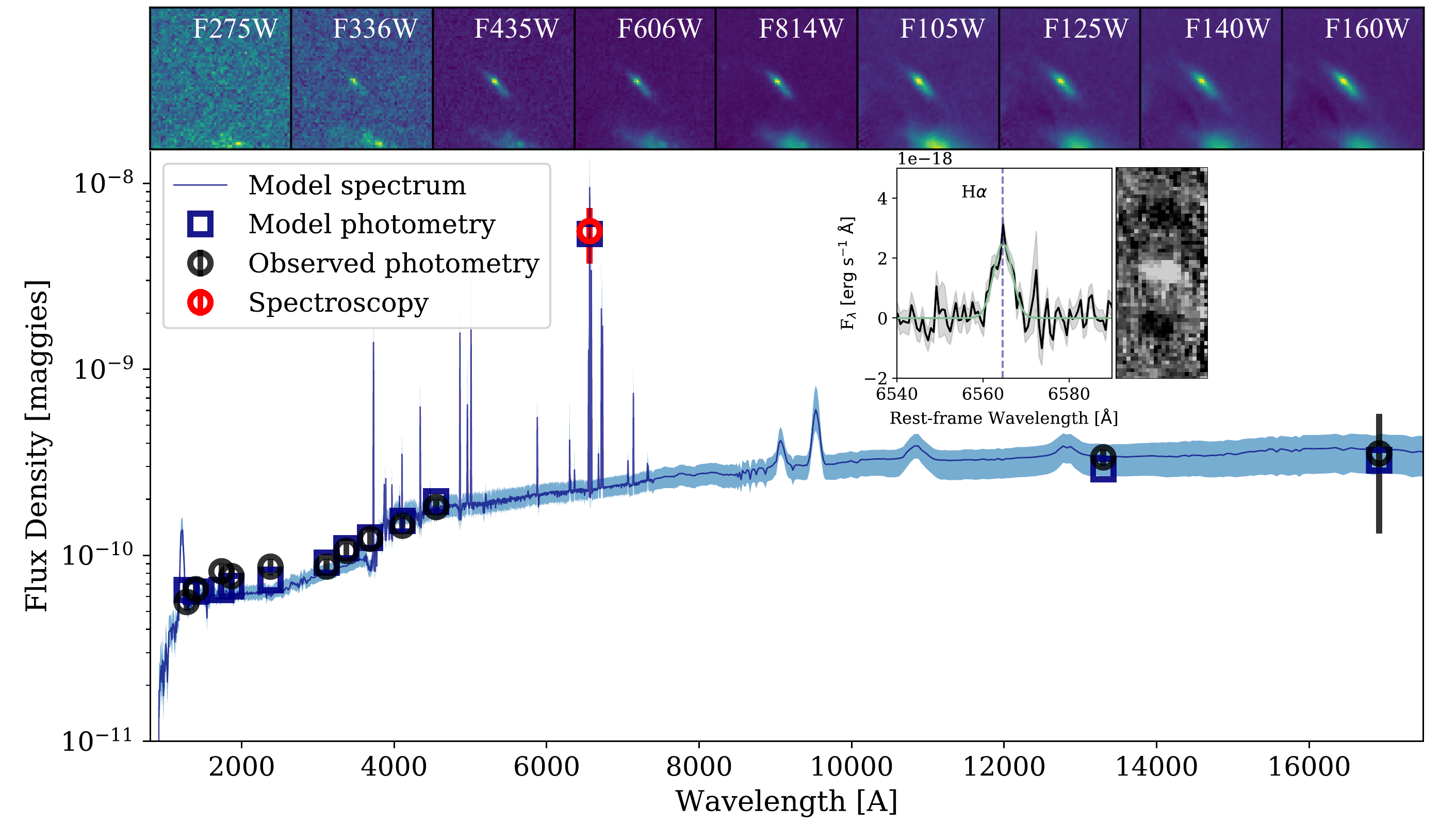}
\caption{The HST images, SED, and MOSFIRE spectrum of the galaxy plus clump. \textbf{Main panel:} \texttt{PROSPECTOR} non-parametric SED fit including constraints from our MOSFIRE H$\alpha$ measurement (shown in orange). The 16th and 84th percentile fits to the SED are shown in light blue. \textbf{Inset panel:} MOSFIRE  \ha \ measurement, where the black line is our raw data, the grey is our error and the teal is our fit to the line. \textbf{Top panels:} non-psf matched photometry in each HST filter considered. 
\label{fig:sed}}
\end{figure*}

Galactic archaeology of the Milky Way (MW) is among the primary testbeds for hierarchical galaxy assembly \citep[e.g.,][]{freeman2002, madau2008, brown2012, bovy2012}.  The ages, abundances, and kinematics of individual stars and star clusters (globular and otherwise) encode the entire formation history of the MW back to the dawn of star formation.  Increasingly detailed studies suggest that our MW had an active accretion and star formation history (SFH) in the early Universe, but has mainly undergone secular evolution in the last several Gyr \citep[e.g.,][]{belokurov2018, haywood2018, helmi2018}.

A number of studies have extended such archaeological techniques beyond the MW. For example, results from the Pan-Andromeda Archaeological Survey (PAndAS) suggest that M31 has had a more active recent accretion history than the MW \citep{mcconnachie2009}.  Similarly, other ambitious efforts with observations of resolved stars and/or integrated light are in the process of revealing the formation histories of MW analogs throughout the Local Volume \citep[e.g.,][]{mouhcine2005, monachesi2016, merritt2016}.

Ideally, it should be possible to connect this galactic archaeology approach directly to observations of accretion and star formation of MW-like ancestors at higher redshifts. Though indirect studies relying on galaxy number densities have shed light on hierarchical formation \citep{leja2013, vandokkum2013}, establishing direct links has proven challenging. This is primarily because even fairly substantial accretion events in the history of the MW \citep[e.g., Gaia-Enceledus,][]{helmi2018} would amount to a 1:10 minor merger, which lies at the mass ratio limit of current high-redshift merger classification schemes \citep{lotz2008, lotz2011, ribeiro2017}. 

At the same time, observations of the Hubble Frontier Fields \citep[HFF;][]{lotz2017} are poised to transform our capacity to study sub-structure (e.g., minor accretion events, star cluster formation) in and around the progenitors of MW-like galaxies at high-redshift.  Compared to blank fields, the magnification power of gravitational lensing in the HFF provides both the spatial resolution and sensitivity to identify and characterize luminous features with sizes and luminosities comparable to star-forming regions and bright satellite galaxies in the local Universe over most of cosmic time \citep[e.g.,][]{swinbank2015, kawamata2015,kawamata2018,Bouwens2017ExtremelyEmissivity,Bouwens2017complexes}.  

Intermediate redshift galaxies (1 < z < 3) have long been known to have higher degrees of sub-structure (usually referred to as ``clumps''
) than their lower-redshift counterparts in restframe optical wavelengths \citep[e.g.,][]{fs2011, Elmegreen2012FORMATIONUNIVERSE,wuyts2012, genzel2011,shibuya2016}. Pioneering studies in the rest-frame optical have even characterized the properties of these clumps as a function of spatial distribution within their host galaxies \citep[e.g.,][]{guo2018, zanella2019}. However, the higher spatial resolution afforded by lensing has shown $\sim$1~kpc clumps to either be multiple more compact clumps, or simply smaller clumps with stellar masses overestimated by up to an order-of-magnitude \citep[e.g.,][]{dz2017, 2018cava}. In principle, such observations can begin to reveal the degree of sub-structure around high-redshift galaxies, allowing us to directly explore hierarchical galaxy formation scenarios in the early Universe.

However, even with improvements in spatial resolution due to lensing, it remains challenging to discern the true nature of these clumps (i.e., \emph{in-situ} star formation vs. accretion).  For example, \citet{zanella2015}, identified an off-center clump in \ot\ Integral Field Unit spectroscopy data, that was not detected in broadband filter observations of a $z = 1.9$ galaxy.  Though the \ot\ implied it was a young cluster, the spatial resolution of the line emission identified clump was $\sim$500~pc, which too coarse for even a star-forming complex \citep[see][and references therein]{krumholz2019}. Additionally, \citet{rujopakarn2019} used ALMA's 30 milliarcsec capabilities to study dust clumping on 200~pc scales in $L_{\star}$ galaxies identified in the Hubble Ultra Deep Field (UDF). However, the resolution of the UDF imaging courser than ALMA, meaning that $\sim~100$pc counterparts in the restframe optical would likely be smoothed out to much larger effective sizes \citep{tamburello2017, gullberg2018}, creating a mismatch between the dust and galaxy scales. Finally, at $z\sim6$, \citet{vanzella2019} combined HFF photometry with MUSE spectroscopy to study Ly-alpha emission from a clump consistent with $r_e < 15$~pc within a 100~pc dwarf galaxy. However, due to intergalactic medium (IGM) absorption, Ly$\alpha$ is of limited use as a star formation rate (SFR) or kinematics indicator. Spectroscopic SFRs (e.g., from H$\alpha$) are necessary to probe star formation within tens of Myr (i.e., the formation timescales of star clusters), but such observations are largely lacking.

In this work, we undertake a joint HFF and Keck spectroscopic study of sub-structure around the progenitor of a MW mass galaxy.  Specifically, we combine K band Keck 1/MOSFIRE spectroscopy with deep HFF photometry, to study an off center ($\sim$100~pc from the galaxy center) $r_e = 40^{+70}_{-30}$~pc, $ M_{\star} \sim10^{8.2} M_{\odot}$ clump within a $r_e = 1.1^{+0.1}_{-0.3}$~kpc, MW progenitor mass ($M_{\star} \sim 10^{9.8} M_{\odot}$) galaxy at $z = 2.36$, with a magnification of $\mu = 5.21$. We jointly model the photomeric SED and \ha\ to determine the SFH of the host galaxy and clump, assuming scenarios in which the 1:40 mass ratio clump is forming \textit{in-situ} and in which it is being accreted. Our joint analysis of broadband photometry and near-IR spectroscopy illustrates the scientific promise of the data that \emph{Hubble} and the \emph{James Webb Space Telescope} (\emph{JWST}) will provide at most redshifts for studying sub-structure, while also highlighting some of the outstanding challenges that must be addressed to increase our understanding of hierarchical galaxy formation at high-redshifts.

\section{Methodology}\label{meth}
We selected targets for spectroscopic follow-up from the HFF MACS J1149, MACS J0717 and Abel 370 clusters, using the foreground subtracted \citet{shipley2018} catalogs with coverage in $F275W$, $F336W$, $F435W$, $F606W$, $F814W$, $F105W$, $F125W$, $F140W$, $F160W$, and IRAC/MIPS from \emph{Sptizer}. 

Initial sizes for our catalog were derived with an updated version of  methodology described in \citet{Bouwens2017ExtremelyEmissivity}.  In short, \com{we mask nearby sources and apply} a \com{lensed} S\'ersic fit for the intrinsic brightness of the source and its size simultaneously using a Markov chain Monte Carlo (MCMC) methodology. As de-lensing an image introduces error and bias depending on the magnification map used, we instead employ a forward modeling approach. Namely, we account for magnification in our models by distorting each pixel by the requisite shear and amplification factors necessary to mimic the effects of lensing. For this to be computationally tractable, we fix the total magnification for our models to the median value from the CATS \citep{cats1, cats2, cats3, cats4}, \citet{bradac}, GLAFIC \citep{glafic1, glafic2, glafic3}, and \citet{johnsonsharon} lensing maps. We discuss the updated size measurement for this source in \S \ref{size}.

When possible, we further selected for sources with photometrically inferred high equivalent width in [OIII] or H$\alpha$, depending on redshift. This was done by running the spectral energy distribution (SED) fitting code, \texttt{FAST} \citep{kriek2009, fast2018}, on our sources and computing the line contamination to the continuum fit from [OIII] or H$\alpha$. Specifically, we fit the photometry excluding the band containing the relevant line to measure the stellar continuum. We then convolve the appropriate filter with the best fit SED and used the difference between this continuum value and the measured photometry to infer line contamination in the band. For sources outside the proper redshift range, we instead selected for high SFRs from the SED fit. \com{This SED fit corresponds to the photometry of the full galaxy, and was only used for selecting sources. We discuss the improved SED fit for this source as well as the SED fits for the clump at the end of this section and in \S \ref{seds}.}

The galaxy we highlight in this work is faint by traditional spectroscopic targeting standards ($F160W = 25$, not correcting for magnification). However, it met our selection criteria for a potentially observable emission line. An early variation of our size measurement routine was biased towards detecting clumps rather than extended sources, which serendipitously resulted in this interesting source falling into our sample. We discuss the full spectroscopic sample in a follow up paper. 

\begin{table}\label{tab:1}
    \centering
    \begin{tabular}{c|c|c}
    \hline
    Parameter & Prior Type & Prior Range \\ [0.5ex] 
    \hline\hline
    Mass/Metallicity & Gaussian  & 7M$_{\odot}<$M$<12.5$M$_{\odot}$/$-1.99<$Log(Z)$<0.0$  \\
    Attenuation ($A_v$) & top-hat & $0<A_v<1.5$\\
    dust index ($\delta$) & top-hat  & $-2.0<\delta<.5$ \\
    Gas Log(Z) & top-hat & $-1.99<$Gas Log(Z)$<0$ \\
    Ionization & top-hat & $-4<$Log(U)$<-1$\\
        
    \hline
    \end{tabular}
    \caption{Summary of priors used to fit the SED for the whole galaxy. We fit mass and metallicity using a joint prior, based off of the \citet{galatzi05} mass/metallicity relation with artificially inflated scatter to account for redshift effects. We use the \citet{kriek2013} dust parameterization and therefore fit for dust index and $A_v$. We also include gas parameters in our model. Though we do not expect our observations to constrain these, including them in our model allows us to treat them as nuisance parameters rather than fixing them to set values. }
    \label{tab:my_label}
\end{table}

The spectroscopic \ha \ detection of this object shown in \autoref{fig:sed} is the result of 6 hours of integration split over two half nights. The reduction was conducted with the \texttt{MOSFIRED} pipeline, an updated version of the reduction procedure described in \citet{kriek2015}. As the source is entirely unresolved, we use an aperture correction for a point source, using a star observed in one of our MOSFIRE slits. No differential aperture correction was required for flux calibration purposes. We then measure a spectroscopic redshift by fitting a Gaussian to the H$\alpha$ line and taking the mean and standard deviation of 10,000 realizations perturbed within the noise as our measurement and error respectively.

We measure the velocity dispersion of this source by fitting a Gaussian to the H$\alpha$ emission line and correcting for the instrument resolution using skylines in the 200\AA\ redwards and bluewards of H$\alpha$. We again derive errors through a Monte Carlo method. 

\begin{table*}
    \centering
    \begin{tabular}{c|c|c|c}
    \hline
    $\mu = 5.14$ & Full Galaxy & Clump Only (ex-situ) & Clump only (in-situ) \\ [0.5ex] 
    \hline\hline
    Redshift & 2.3695 $\pm$ 0.0005 & ---  & --- \\ 
    log(M) [M$_{\odot}$] &  9.8$^{+0.3}_{-0.2}$ & 8.2$^{+0.3}_{-0.2}$ & 8.1$^{+0.2}_{-0.1}$\\
    $A_v$ & 0.44$^{+0.13}_{-0.12}$& 0.15$^{+0.05}_{-0.08}$ & 0.37$^{+0.08}_{-0.02}$ \\
    log(Z) & -0.22$^{+0.30}_{-0.35}$ & -1.35$^{+0.75}_{-0.43}$ & -0.18$^{+0.16}_{-0.23}$ \\
        
    \hline
    \end{tabular}
    \caption{Summary of source and clump properties form our {\sc PROSPECTOR} fits to the photometry from the full galaxy and the clump with ex-situ and in-situ priors}
    \label{tab:my_label}
\end{table*}

\begin{figure}
\includegraphics[width=3.4in]{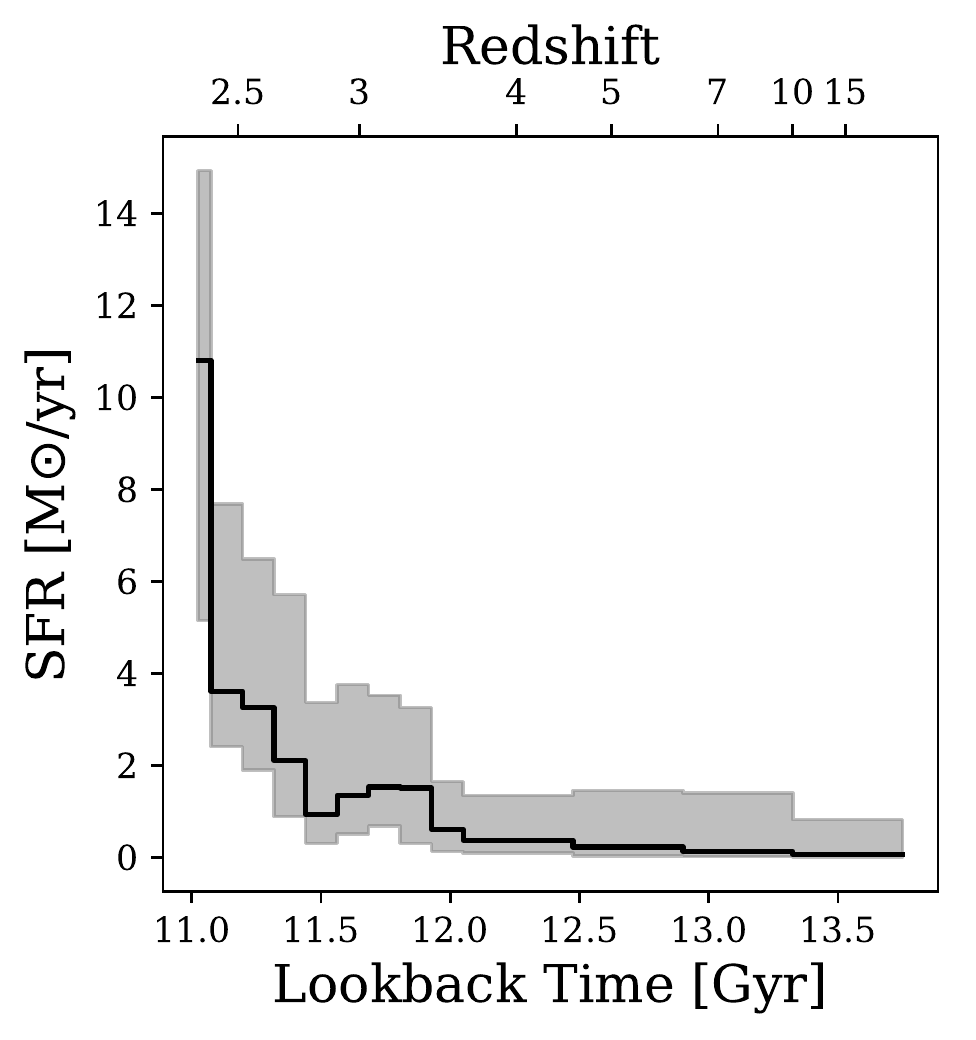} 
\caption{SFH from a non-parametric \texttt{PROSPECTOR} fit. We use 13 age bins in our fit with higher time resolution in the most recent 1 Gyr bins. To avoid biasing our SFH with our choice of age bins, we include more bins than our projected resolution on SFH and apply a continuity prior.
\label{fig:sfh}}
\end{figure}

To measure a non-parametric SFH for our source, we simultaneously fit our MOSFIRE spectrum, and 12 bands of photometry using the Bayesian spectral fitting code \texttt{PROSPECTOR} \citep{leja2017, johnson2019}. To minimize the noise introduced by the lack of continuum detection in our spectrum, we incorporate it into the fit by creating a synthetic narrow band filter over solely H$\alpha$ and including it as an additional band in our photometric fitting, correcting for magnification effects. For the non-parametric fit, we use 13 age bins which are uniformly logarithmically sampled at early times, with finer resolution in the most recent 1 Gyr. To avoid biasing our SFH with our choice of age bins, we include more bins than our projected resolution on SFH, and apply a continuity prior. Following \citet{leja2019_2}, we apply a stellar-mass stellar-metallicity prior using a version of the \citet{galatzi05} relation where we have doubled the width of the confidence interval to accommodate any redshift evolution.

We then use the Bayesian nested sampling code \texttt{dynesty} to sample the posterior \citep{dynesty}. Our most probable parameters are reported in \autoref{tab:my_label}, where errors reflect the 16th and 84th percentiles of the posterior for each parameter, with the uncertainty due to magnification added in quadrature. The magnification uncertainties are modest owing to the small magnification of this source and the geometric constraints afforded by the spectroscopy.

We show the SFH from our \texttt{PROSPECTOR} fit in \autoref{fig:sfh}. We find the star formation is consistent with a rising $\tau$ over a period of a Gyr, with a potential underlying older stellar population at earlier times.  However, the amplitude of the uncertainties preclude a more detailed interpretation of the older SFH.

We quantify the information added by the \ha\ measurement by modeling the SED without it. As expected, the net effect is that the inclusion of  \ha \ provides a tighter constraint on star formation within the most recent 50 Myr, compared to an SED-only fit. Star formation at early times is also modestly more constrained (i.e, the probability distribution function corresponding to bins beyond $1.5$ Gyr is narrower) when our \ha \ measurement is included.

\section{Size Measurements}\label{size}
With a redshift in hand, we are able to improve our estimate of the clump’s size.  To do this, we fix our models to the spectroscopic redshift of the source. This constraint allows us to measure a more precise magnification, \com{decreasing} the original size measured for this object by 5\% over the photometric redshift derived size. 
\begin{figure}
\centering
\includegraphics[width=3.4in]{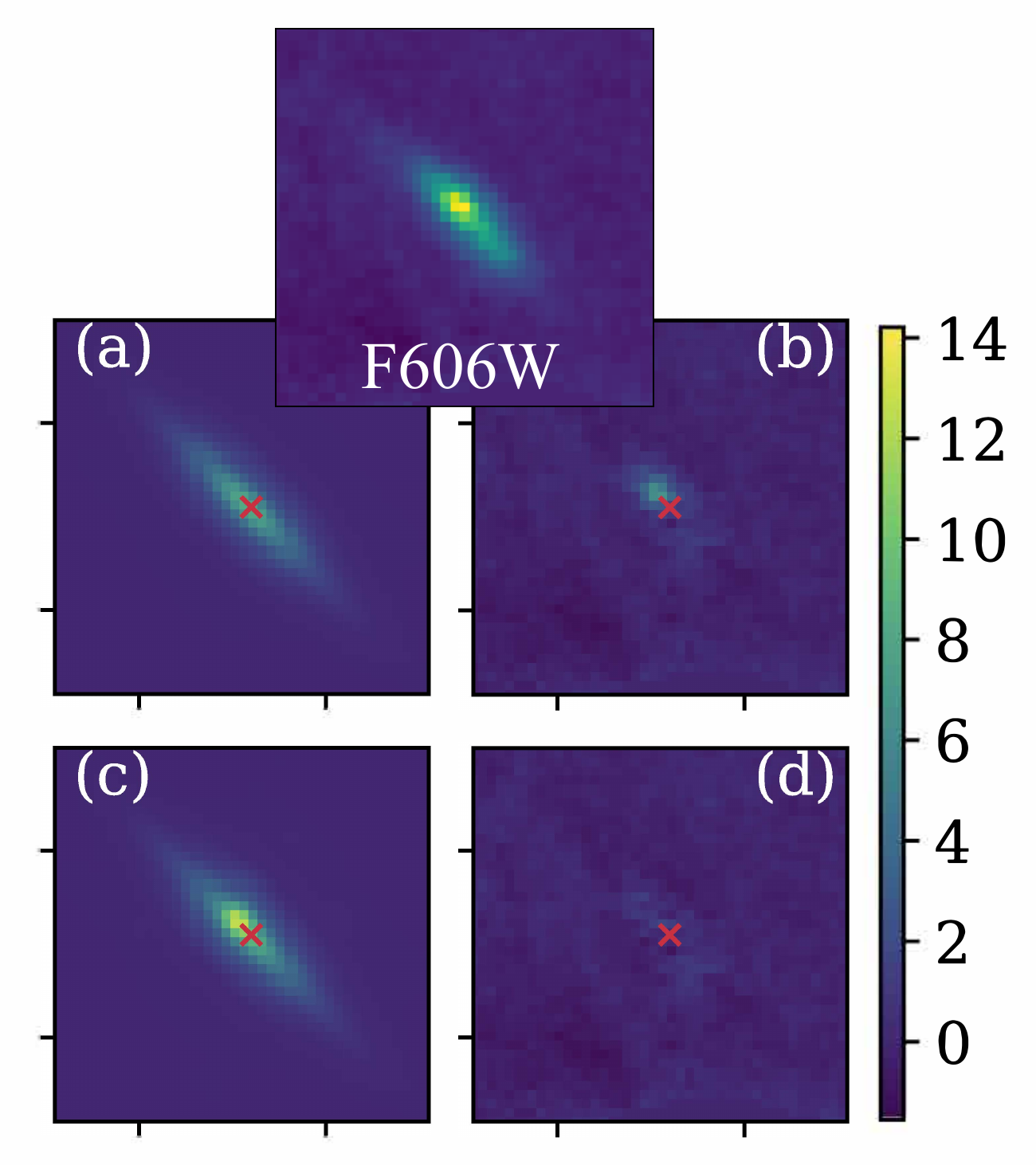} 
\caption{\small Example of size measurement procedure for $F606W$ image. \textbf{(a):} Best fit S\'ersic $n = 1$ model. \textbf{(b):} The residual for the n=1 s\'ersic model with an x demarcating the center of the galaxy accounting for shear. The projected distance from the clump to the center of the source is $\sim100$pc. \textbf{(c):} Best fit Gaussian + S\'ersic $n = 1$ model. \textbf{(d):} Residuals for the Gaussian + S\'ersic $n=1$ model. It is apparent from the last panel that there is an off center clump in the galaxy.     
\label{fig:fits}}
\end{figure}

\begin{figure*}
\includegraphics[width=6.8in]{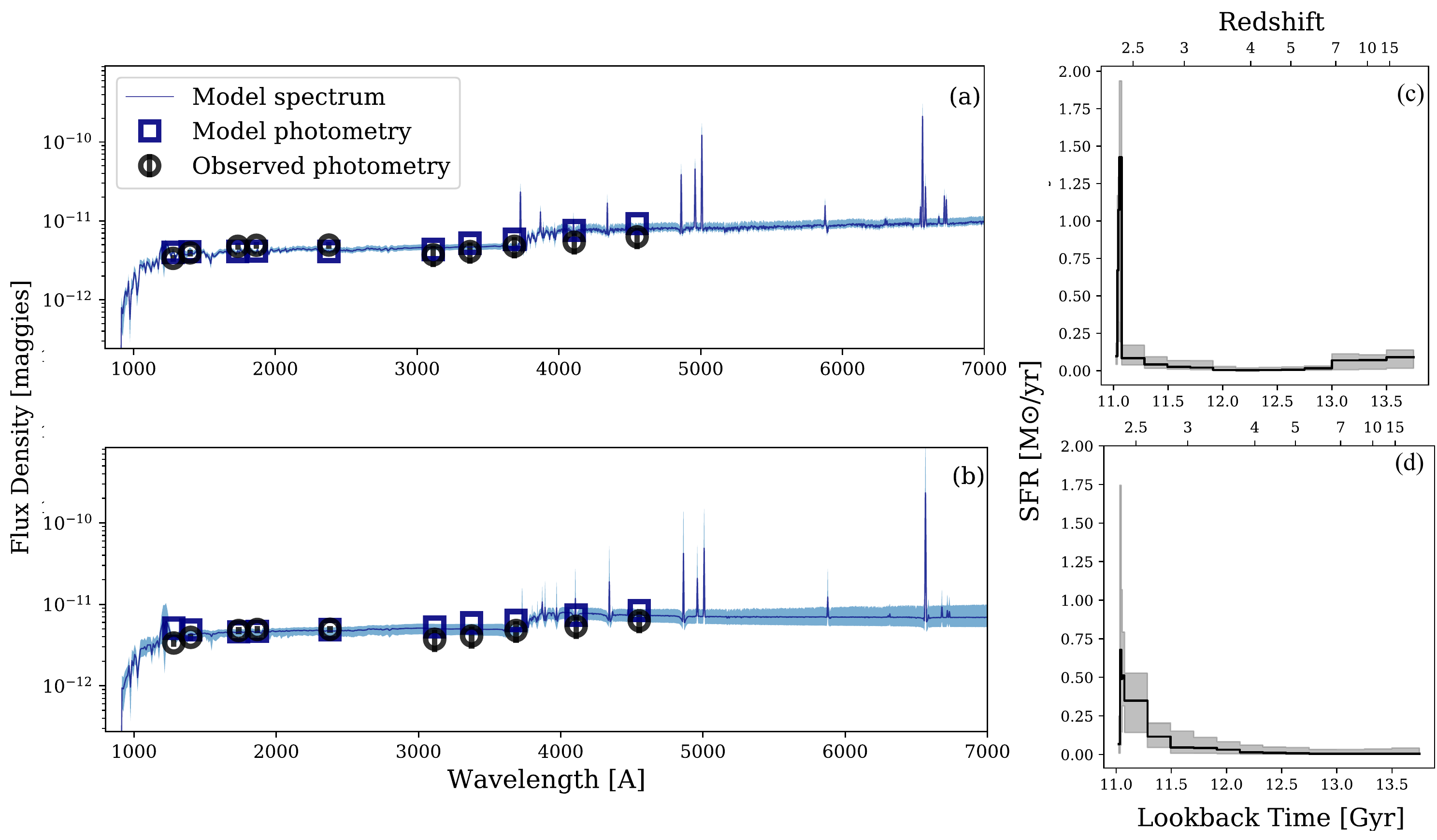}
\caption{The SED fits and SFH for the model extracted clump photometry. \textbf{(a):} The SED fit corresponding to the \emph{in-situ} event priors, with the 16th and 84th percentile models shown in light blue. \textbf{(b):} The SED fit to the accretion formation scenario priors, plotted with the same conventions as panel (a). \textbf{(c):} The SFH corresponding to the \emph{in-situ} scenario. The grey show the 16th and 84th percentiles and the black line shown the median model. \textbf{(d):} The SFH corresponding to the accretion scenario plotted with the same conventions as panel (c). The accretion scenario is a slightly better fit to the data according to the smaller residuals and higher Bayesian evidence criteria. The SFHs resulting from both sets of priors indicate an extremely young stellar population. The burst of the accretion scenario seems to line up with the highest star formation rates in the host galaxy, however, as our first SFH bin corresponds to 50Myr (in concordance with the fit to the host-galaxy), the difference in this small time bin is likely not well constrained by either model.     
\label{fig:mcmc}}
\end{figure*}
\com{To measure the size of the source we began with an unconstrained S\'ersic profile and found that we could not simultaneously fit both the diffuse light component and the clump for any S\'ersic index from 1-4. In an attempt to constrain both components simultaneously, we tested an exponential disk + S\'ersic model, an S\'ersic model + gaussian model, and an exponential disk with a flux-scaled PSF. We find that the latter results in the best residuals by far for all HFF bands, indicating that the clump is essentially unresolved. A conservative upper limit on the size of the clump is therefore the size of the PSF itself, which is 108pc in F606W.}

\com{For a more detailed estimate of the clump size, we fix the S\'ersic profile of the galaxy and simultaneously compute the residuals for a grid of model parameters for a Gaussian clump (size and brightness). We considered values of $r\approx0.02$ pixels to $r=0.5$ pixels for the radius of the Gaussian component. The resulting probability distribution allowed us to constrain the maximum likelihood size of the clump to $r_e  = 40^{+70}_{-30}$pc in F606W.}

\com{Using these model derived parameters, we compare to the background subtracted image within 1$^{\prime\prime}$ aperture, and find the $16~\sigma$ detection shown in panel (b) of \autoref{fig:fits}. We measure the projected distance of the clump from the center of the S\'ersic fit to the diffuse component to be  $100^{+70}_{-30}$~pc. As the center of the S\'ersic fit to the diffuse component is well-constrained, magnification uncertainties dominate our reported uncertainties. This galaxy appears edge-on due to lensing shear, we therefore assume a typical inclination of $45\degree$ to estimate the true distance (d) of the clump to be 140~pc. Comparing this to the effective radius ($R_e$) of the galaxy results in a $d/R_e \sim 0.1$.} 

A caveat to measuring size in bluer bands is that they are biased towards more recent star formation \citep[e.g.,][]{stark2016}. Though we do not expect substantial color gradients in a galalxy of this mass, we asses the robustness of our measured clump by comparing the \emph{maximum a posteriori} value of the MCMC fit for the clump as measured from the $F160W$ to $F606W$ bands. For this clump, the size changes by the resolution difference between $F606W$ and $F160W$. Thus, this source appears more compact in bluer bands primarily due to its point source like nature, rather than bias towards younger populations. As higher redshift studies are limited to probing similarly blue restframe wavelengths in determining sizes, our measured clump size can be directly compared.  

\section{Modeling the SED for In-Situ vs. Accretion Scenarios}\label{seds}
There are two primary mechanisms that can explain the presence of a 1:40 mass ratio for a clump in a proto-MW galaxy; 1: \emph{in-situ} formation due to disk turbulence and gravitational collapse, 2: star formation triggered by an accretion event with a neighboring satellite, which may be analogous to the accretion of an Small Magellanic Cloud progenitor and a MW progenitor at $z\sim2$ \citep{weisz2013}. While the spatial resolution necessary to extract the dynamical information to distinguish these two scenarios will likely remain out of reach until the era of 30-meter telescopes, we can gain some insight as to the likelihood of each scenario by combining spectral and photometric fitting with informed priors.

In the case of \emph{in-situ} disk fragmentation, we expect the metallicity parameters, stellar and nebular, to be similar to the host galaxy parameters. Though the SFH of the clump may differ (possibly dramatically) from the host galaxy, the fact that it is forming out of the galaxy’s cold gas supply implies the metallicity of the clump should be similar to that of its host. \com{To model this scenario, we take the metallicity prior on the clump to be within the 16-84th percentiles of the posterior distribution to the metallicity of the galaxy itself. We leave all other parameters as shown in \autoref{tab:1}}. We assume that though the metallicity of the clump and host galaxy may vary, the metallicity of an \emph{in-situ} clump should not be outside the host galaxy's metallicity range. \com{We then compare this to a 'free' fit to the clump, where the priors are described in \autoref{tab:1}, with the exception of the mass-metallicity prior, which does not apply at such low masses. We therefore instead use separate flat mass and metallicity priors but leave their ranges as shown in \autoref{tab:1}}. The resulting SED fits for the two sets of priors are shown in panels (a) and (b) of \autoref{fig:mcmc}, where the former corresponds to an accretion event and the latter to $in-situ$ formation. 

We find that the model with \textit{ex-situ} priors yields smaller residuals than the \textit{in-situ} priors. However, the Bayesian evidence criteria is of order unity 
which does not signify a robust difference between the two sets of priors. We further show the corresponding SFH for each set of priors in panels (c) and (d). Though these appear qualitatively different, they both essentially correspond to an extremely young stellar population. We discuss future prospects for this archaeological approach to characterizing sub-structure in \S \ref{disc}.

\section{Discussion}\label{disc}

In this section, we explore several plausible clump formation scenarios given the analysis presented in the previous sections.  We attempt to identify features that may allow us to discriminate between accretion and \emph{in-situ} formation mechanisms and discuss how future observations and targeted theoretical studies will improve our ability to cleanly associate observed sub-structure with a formation scenario.

First, we consider the case that this clump formed \textit{in-situ} via turbulence induced disk fragmentation \citep[e.g.,][]{dekel2009, dekel2013}. Cosmological zoom-in simulations have characterized such clumps to have baryonic masses up to $10^9$~M$_{\odot}$ and radii up to 1~kpc. Typical clumps with a baryonic mass corresponding to our measured stellar mass have characteristic radii of $\sim250$pc \citep{oklopcic2017, mandelker2017}. However, as we are limited to measuring stellar mass (we have no gas measurement), our observed clump likely corresponds to a larger baryonic mass in simulations and therefore a larger $R_e$ ($\sim 500$~pc). The mass surface density we measure for the clump is above $10^3$~M$_{\odot} \ pc^{-2}$, which is denser than simulated clumps. Non-cosmological, hydrodynamic simulations described in \citet{tamburellow2015, tamburello2017} do examine the stellar mass and radii of clumps directly as a function of resolution. We find good agreement with our clump mass and radii when comparing to their high-resolution ($100~pc$) mock-observations. 

We can also compare the mass fraction of the clump within the galaxy to simulations. \citet{mandelker2014} find that \emph{ex-situ} clumps have a characteristically higher mass fraction than \emph{in-situ} ones. Comparing to their updated cosmological zoom-in simulations including feedback, we find that our 1:40 mass ratio is on the very massive end of their clump to host mass ratio distribution \citep{mandelker2017}.

The location of our clump, so near the center of its host, remains atypical for clumps formed \emph{in-situ}. Both theoretical \citep{oklopcic2017, mandelker2017} and observational \citep{guo2012, shibuya2016} \com{studies of clump formation find clumps as blue and young as the one we observe further out than $d/R_e \sim 0.1$. The $d/R_e$ value we infer is smaller than the bulk of simulated clumps, but approaches the distance ratio seen in simulations for clumps older than our measured age \citep[e.g., 1~Gyr in][]{mandelker2017} } Finally, disk fragmentation in simulations ubiquitously leads to the formation of multiple clumps \citep{mandelker2014, mandelker2017} and though we have sub-kpc resolution for the entire galaxy, we only identify one. \com{ While, it is possible that there are other clumps that remain obscured due to projection effects, it would require an unlikely geometry to hide all of them from discovery in our data. Given that this clump is detected at $16\sigma$, it is unlikely that we are missing similarly massive clumps due to surface brightness limits.} 

A second \emph{in-situ} formation scenario is that this clump is not a single star-forming entity, but instead an unresolved blend of multiple smaller star clusters. We gain some insight into this possibility by comparing our clump to resolved extreme star-forming regions in the local Universe. For example, the central star-forming cluster of 30 Doradus, NGC 2070, is commonly referred to as an analog for high-redshift star cluster formation \citep{leroy2018, ochsendorf2017}. Integrating the mass of all clusters in a 200~pc region centered on NGC 2070 yields a total stellar mass of $M_{\star} \sim10^5$~M$_{\odot}$ \citep{cignoni2015}, which is 1000 times less massive than our clump.  This is perhaps not surprising given the differences in the host galaxy properties and relative star-forming conditions in the low- and high-redshift Universe.  Nevertheless, it does illustrate just how massive a star-forming region our clump may be.

Perhaps a more apt comparison is to the the merging Antennae system.  The Antennae yields gas pressures analogous to those expected in high-redshift star formation \citep[e.g.][]{cabrera-ziri2016}. Integrating the stellar mass for the three most massive clusters in the Antennae yields a mass of $M_{\star} \sim10^6$~M$_{\odot}$ \citep{johnson2015}. Even assuming the 84th percentile radius for our clump, we would require a mass contamination from the host to be factor of 10 larger than inferred from high-resolution simulations \citep{tamburello2017} or nearby Young Molecular Clouds \citep{2015hollyhead}. Given our high spatial resolution and model extracted photometry, we do not expect anywhere near as severe of a mass contamination. It is therefore unlikely that our clump is a set of very tightly packed massive star clusters.

A third possibility is that an off-center star forming clump may be an indication of an ongoing merger, or pristine-gas accretion event \citep[e.g.,][]{hopkins2012,dsouza2018}. A comparison of the SFH we infer for the observed clump in \autoref{fig:fits} to the SFH of the main galaxy in \autoref{fig:sfh}, shows an extremely young stellar population whose formation corresponds to a peak in star formation in the main galaxy. This concurrence can be interpreted as consistent with an \emph{in-situ} formation scenario or a gas accretion scenario over an accretion event.

However, though both sets of model priors we used for the clump result in a fit consistent with a single burst of star formation, our \emph{ex-situ} priors yield a fit that allows for a more extended ($\sim1$~Gyr) clump SFH. Furthermore, it is difficult to compare our results to the literature on mergers as the redshift and mass range of our source has not been explored at similar resolutions. Morphological criteria, such as $Gini-M_{20}$ \citep{lotz2004} or concentration and asymmetry \citep{conselice2014} have primarily been used at lower redshifts \citep{lotz2011} or higher stellar masses \citep{man2016}. Recent work by \citet{nevin2019}, uses linear decomposition of non-parametric morphological criteria in simulations to find an optimal merger identifying criteria. While work in this vein is promising, it is unclear how well these criteria hold up at higher redshift \citep{2015thompson}. Current criteria for mergers are categorically less sensitive to minor mergers (1:4-1:10 stellar mass ratios) and are insensitive to mergers with larger mass disparity. This means that a MW-Large Magellanic Cloud merger, for example, would likely not be identified with current criteria and a MW-Sagittarius merger would be completely undetectable. With more work in lensed fields, where sub-structure resolution to sub $\sim100$~pc is possible, studying such minor mergers should be within reach.   

Though we cannot conclusively determine the nature of this off-center star forming clump from our current data set, there are promising paths forward.  For example, a larger sample of galaxies and wider spectral coverage with the \emph{JWST}, could enable the modeling technique presented in this work. A statistical approach targeting similarly low magnification sources ($\mu<10$) would allow us to disentangle line-of-site effects that limit our comparisons to simulations (i.e., number of clumps per source). \com{Current explorations of clumps in the HFF generally require observations to approach surface brightness detection limits, with the sensitivity of NIRCAM, it should be possible to detect more clumps than observed in current HFF observations. Likewise, the spectral range, sensitivity and angular resolution of NIRSPEC will allow more efficient spectral analysis of clumps out to higher redshifts, than available with current ground based instruments.}

Current estimates using the \emph{JWST/NIRSPEC} exposure time calculator, show that \ha, \hb, and \ot\ can be detected with $S/N >5$, in under two hours for this galaxy, with faint lines like \ot (4363 \AA ) detected in 5 hours. Such faint lines are easier to measure in more magnified and extreme star forming sources, such as the $z = 2.69$, $\mu \sim 8$, dwarf galaxy identified in \citet{gburek2019}, for which \emph{JWST/NIRSPEC} would detect \ot (4363\AA) with $S/N>5$ in under an hour of exposure time. These lines enable robust dust attenuation measurement and vastly improved constraints on metallicity. These can be used in conjunction with simulations to elucidate the origin of clumps. Furthermore, the multiplexing abilities of \emph{JWST/NIRSPEC} combined with the resolving power afforded by even modestly lensed fields, enable a statistical approach to characterizing clumps in lower mass and higher redshift systems than currently feasible. 

\section{Conclusions}
We are entering an era of increased sensitivity that enables us to study substructure at $z>2$ on ever smaller spatial and mass scales. Our current MOSFIRE study previews the potential of spectroscopy to connect substructure and star formation history, as well as the challenges in interpreting even high spatial resolution photometry. 

In this work, we present a combined spectral and photometric analysis of a lensed, $\mu=5.14$, MW-mass progenitor galaxy at $z = 2.36$ with an off-center star forming clump. We have used precise redshifts to measure robust sizes for the clump and the main galaxy. Finally, we investigate the origin of the clump by leveraging our measurement of the main galaxy to inform priors on the fit to the clump. We find: 

\begin{itemize}
\item A Log(M$_\star$)$ = {8.2}^{+3}_{-2}~$M$_{\odot}$ clump with $r_e = 40^{+70}_{-30}$~pc located $\sim100$pc away from the galaxy center. We find that the host galaxy has a mass of Log(M$_\star$)$= 9.8^{+0.3}_{0.2}~$M$_{\odot}$.  

\item Comparing our inferred clump/galaxy masses, SFRs and sizes to low-redshift analogs and simulations, we find that this clump is unlikely to be an aggregate of multiple less massive clumps. Given our measured stellar mass surface density of order $10^3$ M$_{\odot}pc^{-2}$, we conclude that if this clump formed \emph{in-situ}, it is one of the densest star forming regions confirmed to date. If the clump formed \emph{ex-situ}, this system is undergoing a $1:40$ mass ratio merger.  

\item With the advent of \emph{JWST NIRCAM and NIRSPEC}, we find that this analysis can be reasonably undertaken statistically. The sensitivity and broadband capabilities of \emph{NIRSPEC} will allow measurement of important dust and metallicity indicators. While its angular resolution will allow a more robust comparison between clump and galaxy properties that can be used to infer clump origins. 
\end{itemize}

\section*{Acknowledgements}
We would like to thank the referee for contributing to the clarity and overall quality of the manuscript. This work is based on data and catalog products from HFF-DeepSpace, funded by the National Science Foundation and Space Telescope Science Institute (operated by the Association of Universities for Research in Astronomy, Inc., under NASA contract NAS5-26555). T.O.Z is supported by the University of California Dissertation Fellowship. D.R.W. acknowledges support from an Alfred P. Sloan Fellowship, an Alexander von Humboldt Fellowship, and a Hellman Faculty Fellowship. The authors wish to recognize and acknowledge the very significant cultural role and reverence that the summit of Maunakea has always had within the indigenous Hawaiian community. We are most fortunate to have the opportunity to conduct observations from this mountain.

\bibliographystyle{mnras}
\DeclareRobustCommand{\De}[3]{#3}
\bibliography{mendeley2.bib, references.bib}

\clearpage

\appendix
\section{SED fits and inferred SFH with varying stellar models}\label{appendix}
\begin{figure*}
\includegraphics[width=6.8in]{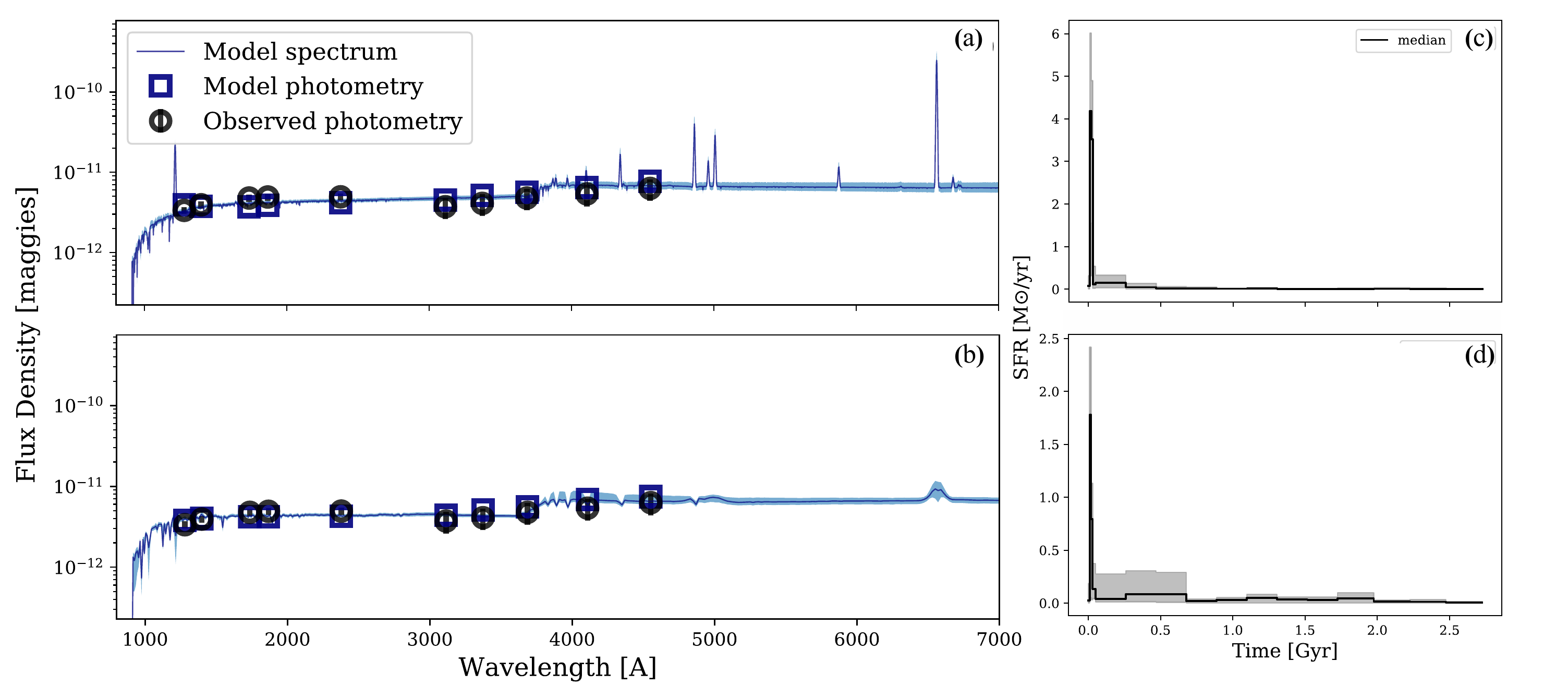}
\caption{The SED fits and SFH for the model extracted clump photometry with alternate stellar libraries. In all plots the shaded regions correspond to the 16th and 84th percentiles. (a): SED fit using the \texttt{BPASS} models with binaries. (b): SED fit using \texttt{BC03} models. (c) SFH corresponding to \texttt{BPASS} model fit. (d) SFH corresponding to the \texttt{BC03} models. Both models show a more peaked median clump SFH, however all models are consistent with a young star forming burst. 
\label{fig:app_fig}}
\end{figure*}

There is currently no standard universally excepted stellar modeling framework in the galaxies literature, potentially leading to systematic offsets in inferred properties across different studies. We investigate alternate isochrone and spectral libraries and their impact on our analysis. We examine both the \texttt{BPASS} models with binaries, as well as the popular \texttt{BC03} which we replicate by combining the \texttt{BASEL} spectral libraries with the \texttt{Geneva} isochrones. We show the results of our \texttt{PROSPECTOR} fits in \autoref{fig:app_fig}. We find that both \texttt{BPASS} and \texttt{BC03} produce a more peaked median SFH than the \texttt{MIST} models (see \autoref{fig:mcmc}). However an extremely young burst of star formation is favored by all three models.

\section{Intrinsic Magnitudes Measured for the Galaxy and Clump }\label{appendix_tab}
\begin{table*}
    \centering
    \begin{tabular}{|c c c|}
    \hline
    Filter & Full Galaxy & Clump  \\
    \hline \hline 
    F336w & $26.67^{+0.35}_{-0.14}$& $28.99^{+1.51}_{-2.50}$ \\
    \hline 
    F435w & $25.63^{+0.33}_{-0.05}$
& $28.81^{+0.67}_{-1.65}$ \\
    \hline 
    F475w & $25.46^{+0.33}_{-0.05}$
& $28.63^{+0.52}_{-1.50}$\\
    \hline 
    F606w & $25.21^{+0.33}_{-0.04}$& $28.33^{+0.41}_{-1.39}$\\
    \hline 
    F625w & $25.28^{+0.33}_{-0.08}$ & $28.34^{+0.49}_{-2.78}$ \\ 
    \hline F814w & $25.15^{+0.33}_{-0.04}$ & $28.35^{+0.49}_{-1.45}$ \\
    \hline F105w & $25.152^{+0.33}_{-0.04}$ & $28.41^{+0.45}_{-1.48}$ \\
    \hline F110w & $24.94^{+0.33}_{-0.05}$ & 28.56$^{0.40}_{-4.38}$ \\ 
    \hline F125w & $24.78^{+0.33}_{-0.04}$ & $28.05^{+0.41}_{-1.39}$ \\ 
    \hline 
    F140w & $24.60^{+0.33}_{-0.04}$& $27.98^{+0.38}_{1.36}$ \\ 
    \hline 
    F160w & $24.35^{+0.33}_{-0.04}$ & $27.95^{+0.43}_{-1.41}$ \\ 
    \hline 
    irac2 & $23.68^{+0.33}_{-0.06}$ & --- \\ 
    \hline 
    irac3 & $23.63^{+0.76}_{-0.68}$ & ---\\
    \hline 
    \hline 
    \end{tabular}
    \caption{Measured intrinsic magnitude for the full galaxy and the clump. The uncertainties on the clump flux are from the model fit and represent the 16th and 84th percentiles of the posterior with the uncertainty in magnification added in quadrature. The uneven uncertainties are due to magnification uncertainties.}
    \label{tab:my_label}
\end{table*}
\clearpage




\bsp	
\label{lastpage}
\end{document}